\def\to{\hbox{$\,$--$\,$}}
\def\muspc{\hskip 0.15 em}

\documentstyle[proceedings]{crckapb} 

\begin{opening}
\title{ARE THE GALACTIC BULGE AND BAR THE SAME ?}
\author{YUEN KEONG NG}
\institute{Padova Astronomical Observatory\\
           Vicolo dell'Osservatorio 5, I-35122 Padova, ITALY}
\end{opening}

\runningtitle{ARE THE GALACTIC BULGE AND BAR THE SAME ?}

\begin{document}

\begin{abstract}
An overview is given of the results obtained about
Galactic Structure studies towards the 
Galactic Centre from star counts. 
The results indicate the presence of an old 
metal-rich population (\mbox{Z\muspc=\muspc0.005\to0.08}; 
\mbox{$t$\muspc=\muspc13\to15~Gyr}), presumably the old 
Galactic Bulge. In addition, a younger, less metal-rich
population was found (\mbox{Z\muspc=\muspc0.005\to0.03};
\mbox{$t$\muspc=\muspc8\to9~Gyr}), likely related to the
tri-axial bar found in other studies.
In Baade's Window a ratio bulge/bar stars  
of approximately 1/2 was obtained from star counts. 
\end{abstract}

\section{Introduction}
The stellar population synthesis technique is used to generate
synthetic Hertzsprung-Russell diagrams (HRDs). 
This is a powerful tool in studies of the properties 
of resolved stellar populations.
The so-called HRD galactic software telescope (HRD-GST)
is developed to study the stellar populations in our Galaxy
(Ng et~al.\ 1995). The basis is formed by the latest
evolutionary tracks calculated by the Padova group
(Bertelli et~al.\ 1994 and references cited therein). 
Through a galactic model synthetic 
Colour-Magnitude diagrams (CMDs) are generated
and compared with the observations.
\par
The primary goal of the HRD-GST is to determine the interstellar
extinction along the line of sight and to obtain constraints on
the galactic structure and on the age-metallicity of the
different stellar populations distinguished in our Galaxy.
Thus far the distribution of the disc stars are described 
by a double exponential with a specific scale height 
and scale length for each subpopulation distinguished
from differences in age and metallicity.
The distribution of the halo/bulge/bar stars are 
well described with a power-law.
One should bear in mind that the CMDs studied
are `snapshots' from our Galaxy. These CMDs do not 
provide information about the Galaxy's dynamical structure.
One therefore cannot distinguish if different stellar populations 
might form one dynamical entity.
The results obtained thus far are reported in various papers 
(Bertelli et~al.\ 1995, 1996; Ng et~al.\ 1995\to1997). 

\section{Bulge = Bar ?}
In star counts studies with the HRD-GST one first has to 
determine the extinction along the line of sight
(Bertelli et al. 1995, Ng{\muspc\&\muspc}Bertelli 1996). 
In one of the fields studied by Bertelli et al. (1995, 1996) 
hot horizontal branch stars were found. 
The Padova isochrones, single stellar populations and synthetic 
stellar populations all indicated that these stars cannot be 
old and metal-poor, but are likely old and metal-rich (13\to15~Gyr,
Z up to 0.08).
Ng et al. (1996) found from a study of the red horizontal branch 
morphology in Baade's Window the presence 
of a significant younger (8\to9~Gyr) and less metal-rich 
(Z\muspc=\muspc0.005\to0.03) stellar population.
The presence of an old, metal-rich population could not
be excluded. However, no conclusive evidence was on the other hand 
found for its presence.
The results thus far indicate that the stellar population constituting 
the galactic bulge and bar are likely not the same. In Baade's Window
a stellar ratio bulge/bar\muspc=\muspc1/2 was obtained.
The star counts studies do not reveal if those
stellar populations behave dynamically like one entity.
This information should be obtained by other means.
\par
Future star counts studies with the HRD-GST ought to 
identify features in CMDs, which are indicative for 
different ages of metal-rich stellar populations. 
Furthermore, studies of large areas are required 
to obtain a statistically significant number of
metal-rich, hot horizontal branch stars.
The extinction ought to be studied carefully in order 
to disentangle metallicity from differential extinction
and a detailed study should be made of the 
(de-reddened) red horizontal branch 
morphology. The availability of the 
OGLE-II data (http://www.astrouw.edu.pl/$\sim$ftp/ogle/index.html) 
satisfies the requirement
for homogeneous photometry from large areas 
and further justifies a forthcoming effort
to verify the star counts results.
\vfill
\medskip
\noindent{\it Acknowledgements}\quad
Ng acknowledges financial support received from the IAU
and the Italian Space Agency (ASI).

\end{document}